\documentclass[preprint]{aastex}

\shorttitle{{\it AKARI} Observation of Early-type Galaxies in A2218}

\shortauthors{Jongwan Ko et al.}

\begin{document}

\title{The Mid-infrared View of Red Sequence Galaxies in Abell 2218 with {\it {\it AKARI}}}

\author{Jongwan Ko\altaffilmark{1,2} , Myungshin Im\altaffilmark{1,2}, Hyung Mok Lee\altaffilmark{1}, Myung Gyoon Lee\altaffilmark{1}, Ros H. Hopwood\altaffilmark{3}, Stephen Serjeant\altaffilmark{3}, Ian Smail\altaffilmark{4}, Ho Seong Hwang\altaffilmark{5}, Narae Hwang\altaffilmark{6}, Hyunjin Shim\altaffilmark{1}, Seong Jin Kim\altaffilmark{1}, Jong Chul Lee\altaffilmark{1}, Sungsoon Lim\altaffilmark{1}, Hyunjong Seo\altaffilmark{1}, Tomotsugu Goto\altaffilmark{7}, Hitoshi Hanami\altaffilmark{8}, Hideo Matsuhara\altaffilmark{9}, Toshinobu Takagi\altaffilmark{9}, and Takehiko Wada\altaffilmark{9}}

\altaffiltext{1}{Astronomy Program, Department of Physics \& Astronomy, FPRD, Seoul National University, Seoul 151-742, Korea}
\altaffiltext{2}{Center of the Exploration of the Origin of the Universe (CEOU), Seoul National University, Seoul, Korea}
\altaffiltext{3}{Department of Physics and Astronomy, Open University, Walton Hall, Milton Keynes MK7 6AA, UK}
\altaffiltext{4}{Institute for Computational Cosmology, Department of Physics, Durham University, South Road, Durham, DH1 3LE, UK}
\altaffiltext{5}{Korea Institute for Advanced Study, Seoul 130-722, Korea}
\altaffiltext{6}{National Astronomical Observatory of Japan, Mitaka, Tokyo 181-8588, Japan}
\altaffiltext{7}{Institute for Astronomy, University of Hawaii 2680 Woodlawn Drive, Honolulu, HI, 96822, USA}
\altaffiltext{8}{Physics Section, Faculty of Humanities and Social Sciences, Iwate University, Morioka 020-8550, Japan}
\altaffiltext{9}{Institute of Space and Astronautical Science, Japan Aerospace Exploration Agency, Kanagawa 229-8510, Japan}

\email{jwko@astro.snu.ac.kr}

\begin{abstract}

 We present the {\it AKARI} InfraRed Camera (IRC) imaging observation of early-type galaxies
 in A2218 at z $\simeq$ 0.175. 
 Mid-infrared (MIR) emission from early-type galaxies traces circumstellar dust emission from AGB stars 
 or/and residual star formation. 
 Including the unique imaging capability at 11 and 15 $\mu$m, our {\it AKARI} data provide an effective way to investigate
 MIR properties of early-type galaxies in the cluster environment.
 Among our flux-limited sample of 22 red sequence early-type galaxies with precise dynamical and line strength measurements
 ($<$ 18 mag at 3 $\mu m$), we find that at least 41\% have MIR-excess emission.
 The $N3-S11$ versus $N3$ (3 and 11 $\mu$m) color-magnitude relation shows
 the expected blue sequence, but the MIR-excess galaxies add a red wing to the relation
 especially at the fainter end. A SED analysis reveals
 that the dust emission from AGB stars is the most likely cause for the MIR-excess, with low 
 level of star formation being the next possible explanation.
 The MIR-excess galaxies show a wide spread of $N3-S11$ colors, implying a significant spread (2--11 Gyr) in the estimated
 mean ages of stellar populations.
 We study the environmental dependence of MIR-excess early-type galaxies over an area
 out to a half virial radius ($\sim$1 Mpc). 
 We find that the MIR-excess early-type galaxies are preferentially located in the outer region.
 From these evidences, we suggest that the fainter, MIR-excess early-type galaxies have just joined the 
 red sequence, possibly due to the infall and subsequent morphological/spectral transformation 
 induced by the cluster environment.

\end{abstract}


\section{INTRODUCTION}

 Early-type galaxies (ETGs hereafter) are the dominant population in galaxy clusters.
 Their stellar population is thought to be homogeneously old and passively evolving 
 as shown in the tight color-magnitude (CM) relation (e.g., Bower et al. 1992; Kodama \& Arimoto 1997).

  This picture of ETGs consisting of homogeneous stellar population
 breaks down when we examine their properties at different wavelengths. It is known from 
 earlier infrared observations that some ETGs exhibit excess far-infrared 
 emission (Knapp et al. 1989), and mid-infrared emission (Knapp et al. 1992; Xilouris et al. 2004).
  The observation in UV also confirms a diversity of ETGs (Smail et al. 1998; Yi et al. 2005; Schawinski et al. 2007).
 A particular interest can be placed on the mid-infrared (MIR) emission.
 Recently, Clemens et al. (2008) found that about 32\% of ETGs in the Coma cluster have
 excess flux over photospheric emission at MIR.

 This MIR-excess can be due to various mechanisms, 
 such as emission from circumstellar dust around
 AGB stars (Knapp et al. 1992), star formation, and AGN activities
 (Quillen et al. 1999). Recent Spitzer observations indicate that the origin of
 the MIR-excess is the dusty AGB stars for the majority of ETGs
 in the red sequence (Bressan et al. 2006).
  According to the model where the MIR-excess is caused by the dust emission from AGB stars,
 the MIR-excess is prominent if the luminosity weighted mean age of the 
 stellar populations is young (Piovan et al. 2003, hereafter P03).

 ETGs with MIR-excess could be used as a tracer of morphological transformation in the cluster environment. 
 For example, star-forming spiral galaxies infalling to a cluster have been suspected that they would be
 morphologically transformed into quiescent S0s through various physical processes
 (e.g., Dressler et al. 1997; Goto et al. 2003; Boselli \& Gavazzi 2006; Park \& Hwang 2008) 
 leaving relatively young ETGs showing MIR-excess.
 However, such an analysis has limited due to the paucity of complete MIR (7 -- 24 $\mu$m) imaging observation 
 covering a large field around a cluster of galaxies.

  In this paper, we examine the relation between the cluster environment and the MIR-excess of
 ETGs in A2218, a galaxy cluster at $z=0.175$ (e.g., Biviano et al. 2004),
 using {\it AKARI} space telescope (Murakami et al. 2007) that offers a wide-field imaging of the galaxy
 cluster at 11 and 15 $\mu$m for the first time. 
 A2218 is a rich cluster (richness class = 4) with a virial radius of (1.5 -- 2.3) Mpc and 
 a mass of (4.8 -- 21)$\times10^{14} M_{\odot}$ (Girardi \& Mezzetti 2001; Pratt et al. 2005).

  A2218 has been observed by the Infrared Space Observatory (ISO),
 and the previous ISO study revealed the lack of MIR-excess early-type
 galaxies at the central region of A2218 (Biviano et al. 2004), while the analysis of optical/NIR imaging and
 spectroscopy of this cluster have shown a wide dispersion in the stellar ages of faint ETGs
 (Smail et al. 2001; Ziegler et al. 2001; Sanchez et al. 2007).
 The two results seem contradictory, and a wide-field IR imaging could offer a solution.

 Throughout this paper, we use \textit{$H_{0}$} = 70 km s$^{-1}$Mpc$^{-1}$,
 \textit{$\Omega_{m}$} = 0.3 and \textit{$\Omega_{\Lambda}$} = 0.7 (Spergel et al. 2003).
 In this cosmology, the angular scale of 1 arcsec at the distance of this galaxy cluster corresponds to 2.97 kpc.
 All magnitudes are given in the AB system.

\section{THE DATA AND THE SAMPLE}

\subsection{Data}

  The rich cluster A2218 is one of the targets selected for studying low redshift clusters
 by the \textit{CLusters of galaxies EVoLution studies} (CLEVL, PI: H. M. Lee; Im et al. 2008), 
 an {\it AKARI} mission program aiming to understand the formation and evolution of galaxies in cluster environments.
 The {\it AKARI} observation for the A2218 field was carried out using 6 broadband 
 filters\footnote{Numbers next to each alphabet indicate central wavelengths}
 (\textit{NIR: N3, N4, and MIR: S7, S11, L15, L24}) of the InfraRed Camera (IRC; Onaka et al. 2007) onboard the {\it AKARI}.
   We covered a 15$\times15$ arcmin$^{2}$ field centered on the X-ray brightness peak of A2218,
 using 4 IRC tiles. The image depth differs according to the locations, and the central 5$\times5$ arcmin$^{2}$ 
 region is the deepest. Table 1 lists the total exposure time and the 5$\sigma$ point-source detection limit
 over 11$^{''}$ diameter aperture in this region. The flux limit at the outer region is
 $\sqrt{2}$ or 2 times shallower than the central region.

 The raw images were processed and stacked with the IRC imaging data reduction pipeline version 070104 
 and then co-added into one final mosaic image using
 SWarp\footnote{$http://terapix.iap.fr/rubrique.php?id\_rubrique=49.$}.
 Cosmic rays were rejected during the stacking.
 Additional removal of cosmic rays was carried out using L.A.Cosmic (van Dokkum 2001) when necessary.
  We checked if this procedure causes flux loss using the North Ecliptic Pole Survey data of {\it AKARI}
 (Matsuhara et al. 2006; Wada et al. 2008; Lee et al. 2009), and found no significant flux loss.

 We used SExtractor (Bertin \& Arnouts 1996) to detect sources.
 The measured ADUs were converted to the Jy unit using the conversion factors from the IRC manual version 1.4.
 When doing the flux calibration for extended sources, we additionally considered color correction 
 since the IRC flux calibration assumed $f_{\lambda}\propto\lambda^{-1}$. 
 We calculated the correction factor following the method provided in the IRC manual with the model SEDs of
 P03.
 We find that the color correction can affect the flux value at $\lesssim 10\%$ level, but $\sim22\%$ flux change 
 occurs in extreme cases. Since it is difficult to know the SED shape a priori, we leave the color correction as
 unknown systematic error.  

 We used the MAG-AUTO to estimate the total magnitude.
 To check the MAG-AUTO as the total magnitude for all band, we estimated large aperture photometry
 of several isolated galaxies in the final image, and found that the difference between MAG-AUTO
 and MAG-APER was within typical measurement errors
 (NIR: $\lesssim$ 3\%, MIR: $\lesssim$ 20\%).
  However, especially in the NIR band, FLUX-AUTO of sources with close neighbors
 are affected by nearby sources.
  To derive fluxes of such objects, we used small circular aperture of 5.$^{\prime\prime}$5 diameter
 and applied aperture corrections that are derived from the growth curve of isolated galaxies with similar flux.

 For the central 10$^{'}\times10^{'}$ region, we also used a deeper L15 image taken for another {\it AKARI}
 program (PI: Serjeant) to derive 15 $\mu$m fluxes. The 5$\sigma$ detection limit of the deep L15 image is 
 20.1 mag, and the details on this data will be presented elsewhere (Hopwood et al. in preparation).

 Additionally, we used optical fluxes ($UBVI$) from Ziegler et al. (2001, hereafter Z01), and 24/70 $\mu$m data 
 taken by the $Spitzer$/MIPS (GTO program $\#$83, PI: G. Rieke) in the archive for the overlapping region with 
 {\it AKARI} where the MIPS 24 $\mu$m image is deeper than the {\it AKARI} image.

\subsection{Sample}

 For the sample selection, we started with 48 ETGs of A2218 of the central 9.7$\times$9.7 arcmin$^{2}$ in Z01
 and additional cluster members whose redshifts were obtained from the NASA/IPAC Extragalactic Database (NED)
 search of the central 15$\times$15 arcmin$^{2}$ area of A2218 (z=0.165 -- 0.185).
 The morphology of 9 sources added from the NED search is determined to
be ETGs by the visual inspection of the archival optical images (CFHT and Subaru) and SED shapes.
These galaxies were matched with the {\it AKARI} source catalog, and we checked each member whether it is blended or not in the {\it AKARI} image by comparing with the optical images.
Through this process, we excluded galaxies that are blended with neighbors in the {\it AKARI} image.
 Finally, we collected 39 (30 from Z01) non-blended ETGs from the {\it AKARI} data. 
 Among these we have selected a sample of 22 galaxies with $N3 < 18$ mag for MIR-excess study.

\section{RESULTS \& DISCUSSION}

\subsection{The color-magnitude relation}

  In Figure 1, we show the CM diagram of A2218 galaxies in three colors; $U-V$
  (a),  $N3-N4$ (b), and $N3-S11$ (c). 
  The $N3$ flux is chosen to be on the x-axis, as a rough measure of the 
  stellar mass (e.g. Gavazzi et al. 1996). 
  The $U-V$ versus $N3$ CM diagram is shown for galaxies with $N3 < 18$ mag. 
  It shows a usual tight red sequence of ETGs,
  with two galaxies ($\#665$ and $\#2139$ in Z01) slightly bluer than the others.
  This reflects the fact that Z01 sample is made of galaxies belonging to
  the red sequence or close to it.
  Figure 1b shows the $N3-N4$ versus $N3$ CM relation. The plot shows a tight  
  blue sequence since the flux decreases as the wavelength increases at $\lambda > 1.6$ $\mu$m. 
  Next, we have examined the MIR ($S11$) properties of galaxies selected using a box shown in Figure 1b,
  i.e., galaxies with $N3 < 18$ mag.
  This magnitude cut is imposed due to the relatively shallower $S11$ detection limit, so that 
  we can construct an unbiased sample in $N3-S11$ colors.

  Figure 1c shows the $N3-S11$ versus $N3$ CM relation of the cluster members with $N3 < 18$ mag.
  Also plotted are the CM relations from various model SEDs (see below).
  At the bright end, four galaxies tend to form a blue sequence indicating that they are a homogeneous population. 
  However, Figure 1c reveals a stunningly wide dispersion in the $N3-S11$ colors of 
  the optical red sequence galaxies with $N3 > 17$ mag. More than $\sim$50\% show MIR-excess at $N3 > 17$ mag 
  above the model line considering photospheric emission 
  only (the ``old noAGB'' line). 
  This is in clear contrast to the result
  of Biviano et al. (2004) where a small fraction of ETGs showed this kind of MIR-excess.

 To understand what is causing the MIR-excess,
 we compared the observed CM relation with the model predictions. For the model SEDs, we used 
 Single Stellar Population (SSP) models with a spontaneous burst, the Salpeter (1955) IMF,
 and the Padova Library of stellar spectra from P03 which also incorporates the dust emission
 from circumstellar dust around AGB stars.
 As it has been recognized, the optical CM relation can be described well with 
 a single age model assuming a metallicity gradient (the dashed line in Fig. 1a; e.g., Kodama 
 et al. 1997). The same model fits the $N3-S11$ versus $N3$ CM relation at the bright magnitude (the solid line
 in Fig. 1c), but it fails to reproduce the dispersion in the $N3-S11$ colors of the fainter ETGs. 
 The dispersion requires the existence of stellar populations with younger ages, or
 some other mechanisms.

 In the top panel of Figure 2, we examine the MIR-excess of ETGs
 in more detail by plotting $N3-S11$ versus $N3-N4$ color-color diagram. 
 The thick solid line indicates the track of a SSP as described in 
 the dashed line of Figure 1a but with several different ages (1$\sim$15 Gyrs). 
 The other lines indicate the tracks of model SEDs with the AGB circumstellar dust of P03.
 We define MIR-excess galaxies to be those lying at the redward of the dotted line.
 This corresponds to the MIR-excess more than typical 1$\sigma$ error of ($N3-S11$) color ($\sim0.28$ mag)
 over the thick solid line. 
 Among galaxies in our magnitude limited sample, more than 41\% (9/22) 
 can be classified as galaxies having MIR-excess.
 In the lower panels of Figure 2, MIR-excess galaxies have relatively low velocity dispersion.
 Although considering our flux-limited sample, this result suggests that less massive ETGs 
 are more likely to have MIR-excess than massive ones. 
 However, MIR-excess does not show strong correlation with H$\beta$ line index which is sensitive to recent
 star formation activity. 
 It is very possible that H$\beta$ line and MIR-excess trace different epoch in star formation history,
 but it seems difficult to draw a firm conclusion on this based on the current data alone.

  Figure 3 shows the observed optical-MIR SEDs of nine $S11$-detected galaxies.
  We performed the least square fit of SEDs using three different models -- 
  spectral synthesis models with or without AGB dust from P03 ($AGB$ or $noAGB$ model),
  and templates of star-forming galaxies from Chary \& Elbaz (2001). 
  The best-fit SEDs from each model are overplotted in Figure 3.
  The SSP models of P03 have three different metallicities ($Z=0.02, 0.008, 0.004$) and 
  a wide range of ages (0.1$\sim$15 Gyrs).
  In case of the fit to the SF model, we do not use the optical data for the fit,
  since the empirical SF SED templates do not include the diverse optical SEDs with 
  different ages, metallicities, and star formation histories.

  Based on the significance of MIR-excess, 
  we categorize the sample into three sub-groups; strong-, mid-, and weak/no MIR-excess galaxies.
  First, we note that the P03 $noAGB$ model fails to fit the strong/mid MIR-excess galaxies, but provides 
  reasonable fits to the weak/no ones.

  Next, we find that most ETGs with MIR-excess returned acceptable SED fits to the P03 $AGB$ model 
  with either a solar or 40\% solar metallicities.
  The lowest metallicity P03 model (20\% solar) can fit
 the MIR-excess if we consider the IR portion of the SEDs alone, but such a model fails to fit the optical
 portion of the SEDs.
  The results suggest that the MIR-excess correlates with ages more strongly than 
 metallicities, with a distinct trend that the younger galaxies have the stronger MIR-excess 
 (Temi et al. 2005).
  For the MIR-excess ETGs, the derived mean stellar ages 
 span a wide range of 2 to 11 Gyr, which qualitatively agrees with the findings
 by Smail et al. (2001) and Biviano et al. (2004).
 This suggests that the MIR-excess is a good indicator of mean stellar ages of ETGs, 
 as also noted by Temi et al. (2005).

  The AGB dust emission is not the only possible cause for the MIR-excess. We caution that 
 the MIR-excess may arise from the star-forming activities and AGNs. In some cases, it is difficult to exclude 
 such possibilities from the current data alone. 
 Deeper 70 $\mu$m data could offer an effective way to test these possibilities. 
 However, even if the MIR-excess arises from the star-forming activity, 
 the derived SFRs from the SED fits suggest SFR $< 1 M_{\odot}$ yr $^{-1}$ for all of the MIR-excess galaxies. 
 The only exception is $\# 697$ in Z01 which shows an abnormally high 15 $\mu$m flux.
 It is difficult to explain such a flux with the P03 $AGB$ model.
 This may be a piece of evidence for existence of PAH emissions from star formation for this galaxy. 
 The morphology of this galaxy seems to show a disk. The objects $\#1914$, and $\#2139$ have spiky 11 $\mu$m fluxes, 
 which may be due to unusual PAH features at 11.3 and 12.7 $\mu$m found in some ETGs 
 (e.g. Kaneda et al. 2005).

 We conclude that more than half of cluster ETGs with $N3 <$ 18 mag are likely to have
 the MIR-excess over stellar photospheric emission, which we interpret as 
 the emission from circumstellar dust around AGB stars.
 Our result is different from the ISOCAM study of Biviano et al. (2004)
 where they did not find such MIR-excess for most of ETGs in A2218.
 This can be attributed partly to the deeper depth of our data, but mostly to our much wider area coverage.

\subsection{Environmental dependence of MIR-excess}

  To investigate the MIR properties of A2218 ETGs as a function of the cluster environment,
  we plot the spatial and velocity distributions of 39 {\it AKARI} cluster ETGs with the contour maps of the 
  $Chandra$ X-ray emission in Figure 4.
  MIR-excess ETGs seem not to have a preference for the velocity distribution, but the clustercentric-distance
  distribution of them is exceptional.

  We expect to find younger, MIR-excess galaxies in the outskirts
  of the cluster if such galaxies are formed via transformation from field spirals accreted to
  the cluster environment. To check this, we divide our samples into those closer to the center and 
  those in the outer region depending on the cluster-centric distance (at 381 kpc) 
  in the same manner of Z01.

  We find that about 64\% (7/11) of the outer galaxies have MIR-excess, while only 18\% (2/11) of the inner
  ones do so. These MIR-excess galaxies are also relatively faint ones in our sample. 
  We remark that there are ETGs in the inner region as faint as the MIR-excess ETGs
  in the outer region, but they do not show clear MIR-excess.
  From these results, we conclude that these MIR-excess galaxies are fainter,
  and preferentially located in the outer region of the cluster, suggesting that they could be the descendants of 
  the infalling field galaxies transformed into ETGs.
  Our conclusion appear to be reconciled with the result of Z01 where
  they found larger spread of H$\beta$ values of A2218 cluster ETGs in the outskirts.
  We are carrying out similar studies using 6 more nearby clusters to obtain a more conclusive answer.

\acknowledgments
 We thank L. Piovan for providing his SED model. We also thank the referee, Bodo Ziegler, for
 his constructive comments which improved our paper.
 This work is based on observations with {\it AKARI}, a JAXA project with the participation of ESA.
 We acknowledge the Creative Research Initiatives program, CEOU of MEST/KOSEF. 
 RHH, SBGS, and IRS also acknowledge support from STFC.

\clearpage

\begin{deluxetable}{cccc}
\tablenum{1} \tablecolumns{4} \tablecaption{Observational parameters of the A2218 data  \label{tbl-1}} \tablewidth{0pt}

\tablehead{Filter name & Effective wavelength & Total Integration time & 5$\sigma$ point-source sensitivity  \\
\colhead{} & [$\mu$m] & [second] & (within 11$^{\prime\prime}$ ap.) [AB mag.] }

\startdata

N3  & 3.2  & 710.6 & 21.3  \\
N4  & 4.1  & 710.6 & 21.5  \\
S7  & 7.2  & 392.7 & 20.0  \\
S11 & 10.4 & 490.9 & 19.6  \\
L15 & 15.9 & 785.4 & 19.3  \\
L24 & 23.0 & 785.4 & 18.3  \\

\enddata

\end{deluxetable}

\clearpage

\begin{figure}
\epsscale{.80}
\plotone{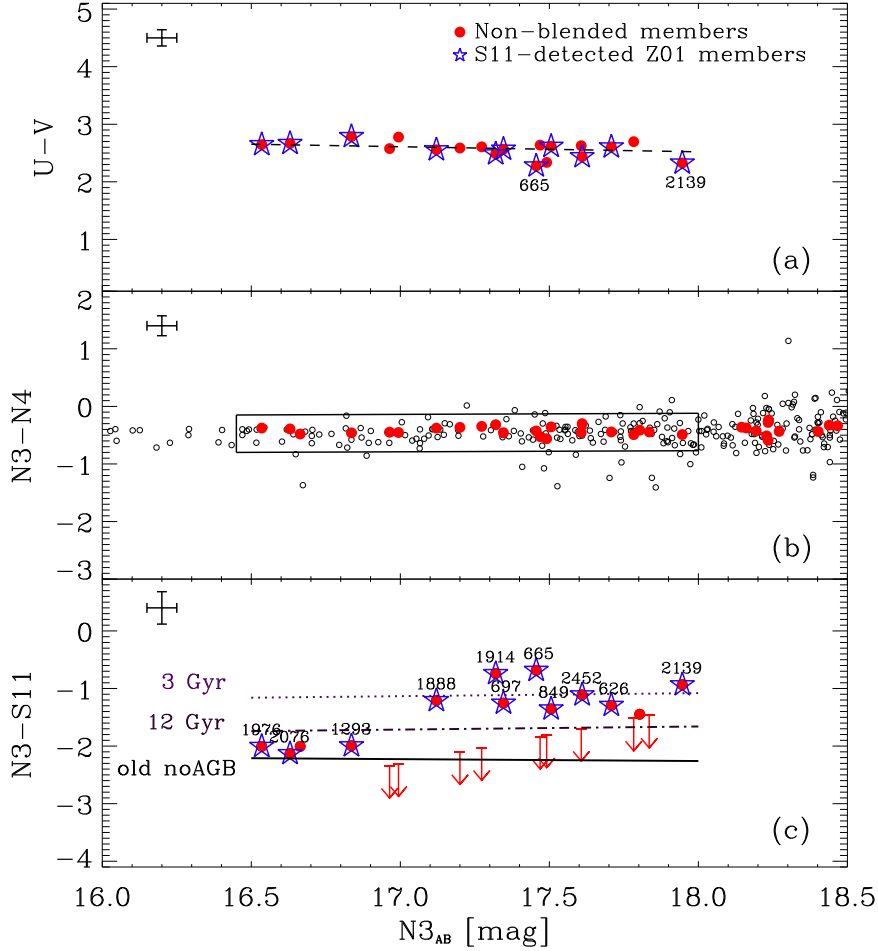}
\caption{\textit{Top}: The $U-V$ vs. $N3$ CM diagram for N3 flux limited
 ETGs. Filled circles indicate our non-blended ETGs, 
 and among them Z01 members detected in $S11$ are represented by star symbols. 
 The galaxy numbers ($\#665$ and $\#2139$) are the notation used by Z01. 
 The dashed line represents the best-fit $U-V$ vs. $N3$ CM relation.
 The cross in the upper left corner indicates the median errors.
 \textit{Middle}: The $N3-N4$ vs. $N3$ CM diagram of {\it AKARI} IRC imaging sample.
 Open circles are all sources detected in the 15$\times$15 arcmin$^{2}$ field.
 Due to the detection limit of the $S11$ flux,
 we select bright ($N3 <18$) sources.
  \textit{Bottom}: The $N3-S11$ vs. $N3$ CM diagram for the galaxies in the box 
  of panel (b).
 The 3$\sigma$ detection limit are given for sources undetected in $S11$ (arrows).
 The dotted and the dashed-dotted lines indicate the CM relation
 calculated from the P03 $AGB$ model SEDs assuming the metallicity gradient
 at two different stellar ages (3 and 12 Gyr), respectively. The solid line
 represents the 12 Gyr P03 $noAGB$ model, assuming the metallicity 
 sequence that fits the $U-V$ vs. $N3$ CM relation. \label{fig1}}
\end{figure}

\clearpage

\begin{figure}
\epsscale{.80}
\plotone{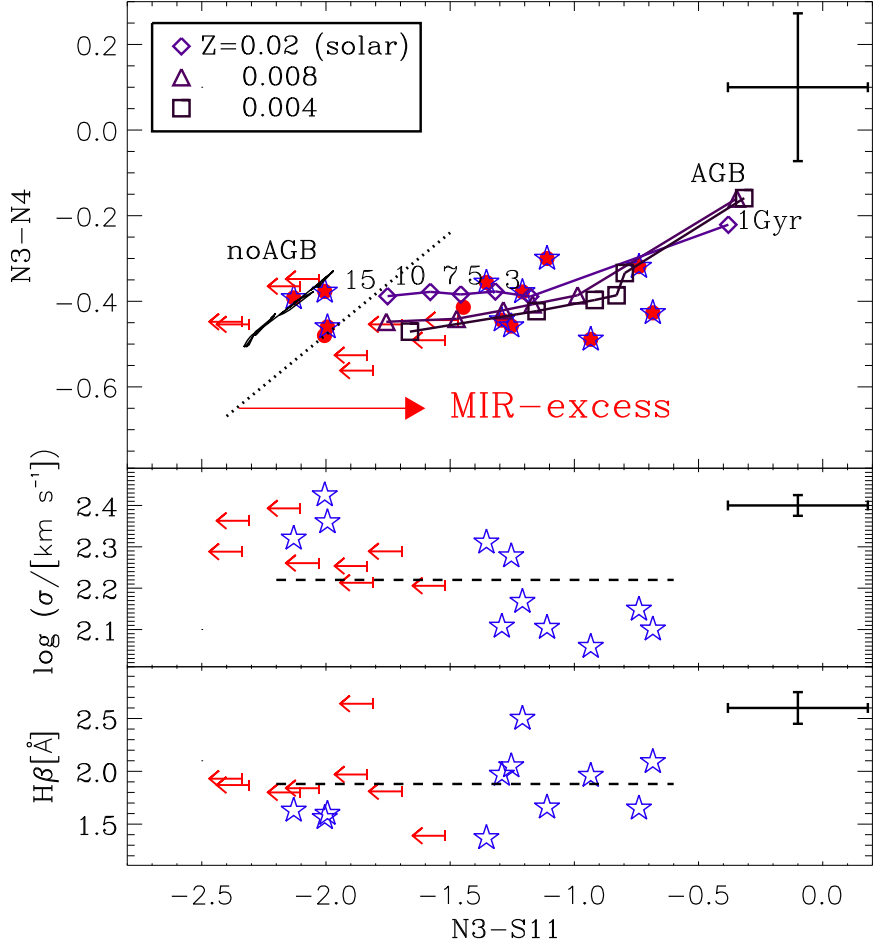}
\caption{\textit{Top}: The $N3-N4$ vs. $N3-S11$ color-color diagram for galaxies 
 in the box of Fig. 1b. Symbols have the same meanings as Fig. 1c.
 The diamond, triangle, and square symbols represent loci of the P03 $AGB$ models
 along mean stellar ages (1, 3, 5, 7, 10, and 15 Gyr) for Z=0.02 (solar metallicity), 0.008, and 0.004, respectively.
 The thick solid line shows the P03 $noAGB$ model color.
 $MIR-excess$ galaxies as those redder than dotted line in N3-S11 (1$\sigma$ redder than the P03 $noAGB$ model). 
 The cross represents the median errors.
 \textit{Middle}: The velocity dispersion vs. $N3-S11$ color for Z01 data available.
  Symbols as the upper panel.
  The dashed line represents the mean value of $\sigma$ for 48 ETGs in Z01. 
 \textit{Bottom}: H$\beta$ line index vs. $N3-S11$ color for Z01 data available.
  Symbols as the upper panel. The dashed line represents the mean value of H$\beta$ line indices for 48 ETGs in Z01. 
 \label{fig2}}
\end{figure}

\clearpage

\begin{figure}
\epsscale{.90}
\plotone{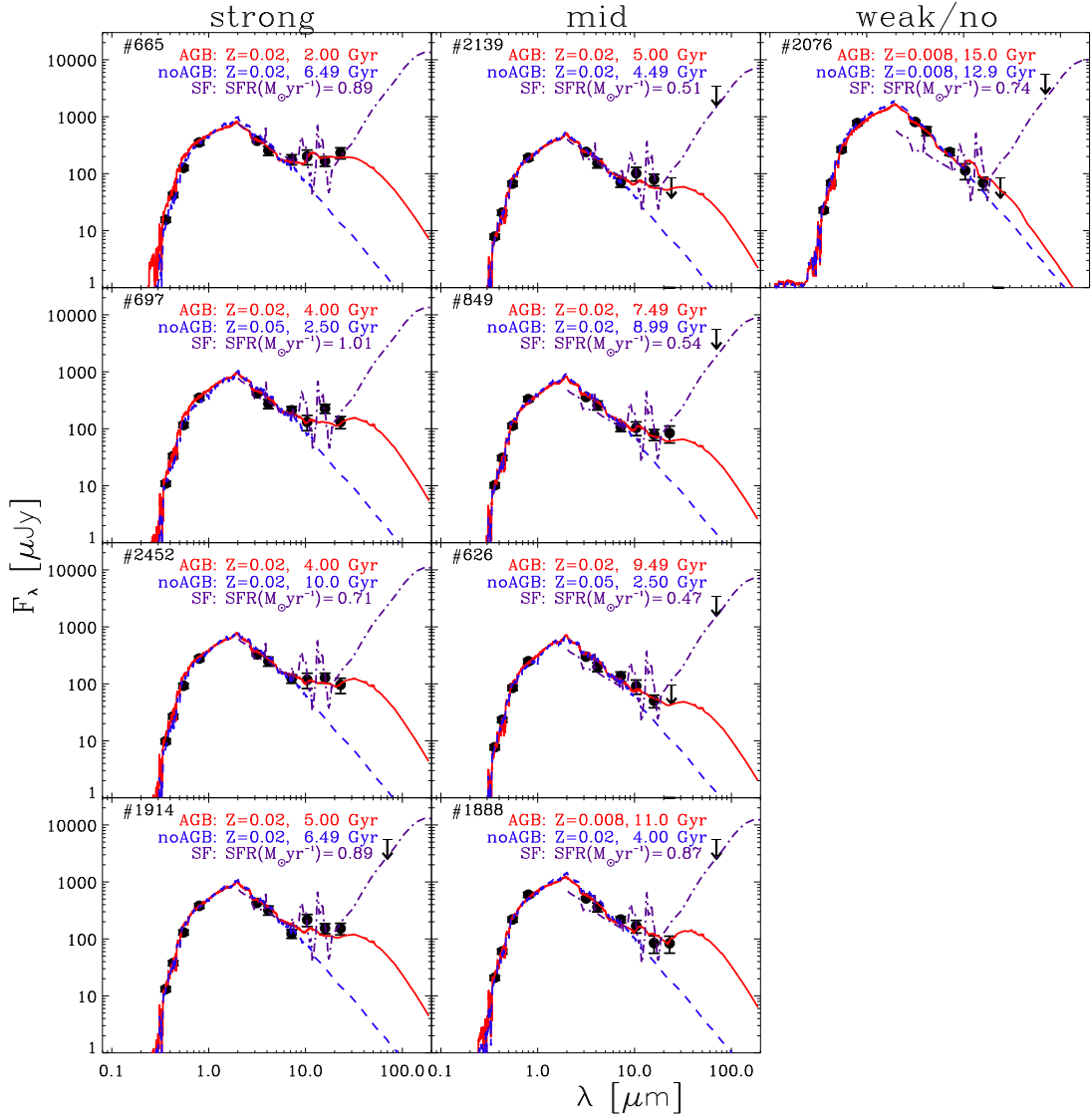}
\caption{The observed SEDs of the $S11$-detected galaxies with Z01 optical and the $Spitzer$ 24/70 $\mu$m 
 data available. The best-fit lines of three different models ($AGB$: solid, $noAGB$: dashed, $SF$: dashed-dotted)
 are overplotted with 1$\sigma$ error bars.
 Arrows represent the 3$\sigma$ detection limit.\label{fig3}}
\end{figure}

\clearpage

\clearpage
\begin{figure} 
\epsscale{.90}
\plotone{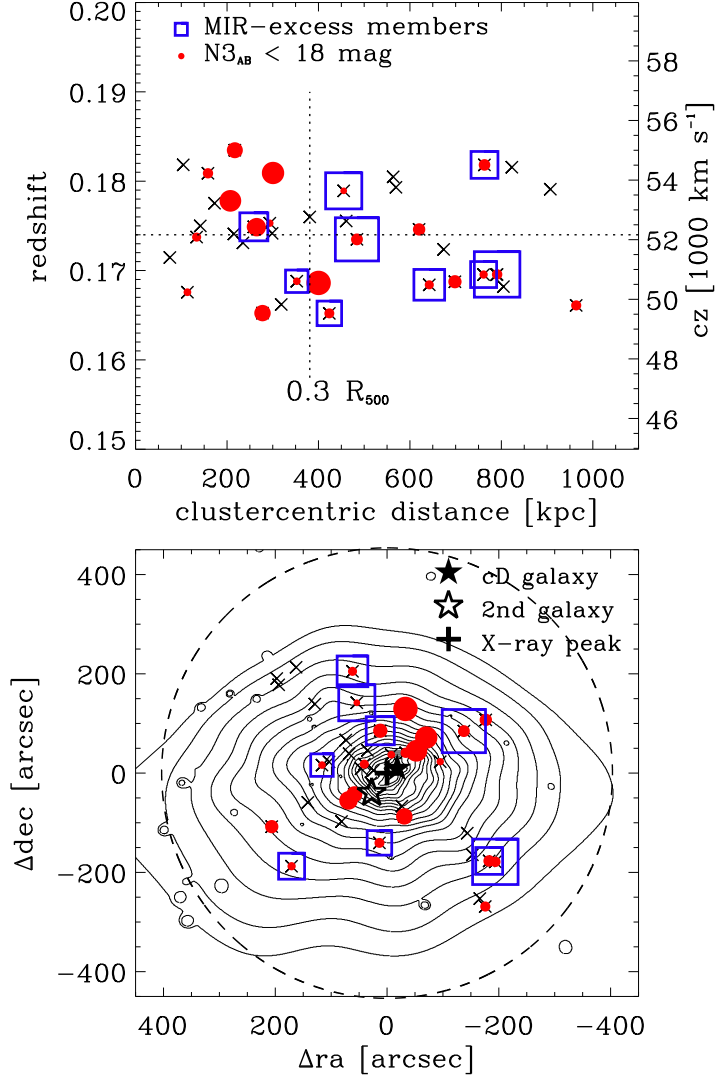}
\caption{\textit{Upper}: The clustercentric distance and velocity distributions for the 39 A2218 {\it AKARI} cluster ETGs
 ('$\times$' symbols). Filled circles are ETGs with $N3 < 18$ mag, and the size is proportional to $N3$ flux. 
 The squares are MIR-excess galaxies as defined in Fig. 2, and the size is proportional to 
 $F_{11}/F_{3}$ flux density ratios (mass-normalized MIR-excess). The vertical dotted line indicates the radius to 
 divide into two bins and $R_{500}$ is usually known as roughly half of the virial radius 
 (1.21 Mpc from Maughan et al. 2008).
  \textit{Lower}: The spatial distribution overlaid with the adaptively-smoothed X-ray contours
   adopted from Maughan et al. Symbols as the upper panel.
   The cross, the filled star, and the open star indicate the position of the X-ray brightness peak, the cD galaxy, 
   and the 2nd brightest cluster galaxy, respectively.
   The dashed circle corresponds to the radius $R_{500}$.\label{fig4}}
\end{figure}

\end{document}